# A Model of the Mechanisms Underlying Exploratory Behaviour

**Liane M. Gabora and Patrick Colgan**

**ABSTRACT**

A model of the mechanisms underlying exploratory behaviour, based on empirical research and refined using a computer simulation, is presented. The behaviour of killifish from two lakes, one with killifish predators and one without, was compared in the laboratory. Plotting average activity in a novel environment versus time resulted in an inverted-U-shaped curve for both groups; however, the curve for killifish from the lake without predators was (1) steeper, (2) reached a peak value earlier, (S) reached a higher peak value, and (4) subsumed less area than the curve for killifish from the lake with predators. We hypothesize that the shape of the exploration curve reflects a competition between motivational subsystems that excite and inhibit exploratory behaviour in a way that is tuned to match the affordance probabilities of the animal's environment. A computer implementation of this model produced curves which differed along the same four dimensions as differentiate the two killifish curves. All four differences were reproduced in the model by tuning a single parameter: the time-dependent component of the decay-rate of the exploration-inhibiting subsystem.

## 1. INTRODUCTION

Selection tends to favour the evolution of systems whose organization enables more efficient ways of perceiving and interacting with the environment, and greater capacity to cope flexibly with environmental change. This often entails progressive differentiation of the parts of a system into subsystems that are specialized to take care of different aspects of survival. For example, the parts of an animal involved in the detection and avoidance of predators can be thought of as comprising one subsystem, the parts involved in obtaining and metabolizing energy another, and so on.

Thus, an adaptive system can be thought of as a set of specialized subsystems working together to maintain the integrity of the whole. However, when we divide the system into subsystems according to ultimate causal goals such as avoiding predators and obtaining energy, every subsystem that relies on behaviour includes skeletomusculature. Much of the skeletornusculature plays a role in the functioning of many subsystems. This makes sense; subsystem-specific limbs are redundant, to the extent that (1) subsystems require only periodic control of skeletomusculature to function effectively, and (2) the skeletomusculature required to meet the needs of one subsystem could also be used to meet the needs of another. The upshot: skeletomusculature is shared amongst subsystems, and though subsystems have to work together cooperatively, they must also arr./Me for control over what McFarland and Sibly (1975) refer to as the system's "behavioural final common path" (see for example, Miller, 1971; Baerends and Drent, 1982; Colgan, 1989). Which subsystem wins the competition depends upon the relative need (deviation from homeostasis) of the subsystems, the opportunities and dangers currently afforded by its environment, and the pros and cons of engaging in behaviour that has only indirect or long-term effects. Dawkins (1976) has suggested that subsystem competition is lessened somewhat by the fact that the behaviours associated with various subsystems occupy different positions on an established behavioural hierarchy; given equal need to control behaviour, the behaviour that is higher on this hierarchy wins the competition. The primary purpose of this research was to explore the extent to which we can explain and predict behaviour in a novel environment by viewing it as the emergent outcome of the continual process of *mutually satisfying the constraints imposed by competing, specialized subsystems*[1]. An early theory of behavioural response to novelty, states that the extent to which an animal explores its environment depends on the outcome of a competition between the tendency to avoid novel stimuli because they evoke fear, and the curiosity-induced tendency to approach (Montgomery, 1955). Subsequent research has shown that when animals encounter a novel environment, exploratory behaviour increases, reaches a maximum, and then decreases; thus, plotting exploration against time yields an inverted-U-shaped curve (Weisler & McCall, 1976). This finding has held across a variety of environmental stimuli and organisms, including humans (May, 1963; Karmel, 1969; McCall, 1971, 1974), chimpanzees (Welker, 1956a, 1956b), monkeys (Mason, 1961; Sackett, 1972), and rats (Derriber and Earl, 1957). However, there has been no attempt to employ the Curiosity-Fear theory to account for the inverted-U-shaped curve; in fact little attention has been given to the mechanisms that underlie the phenomenon.

We hypothesize that in a well-adapted animal, a number of subsystems simultaneously excite or contribute to exploration of the environment, including those that can benefit from the finding of food, mates, or shelter, or from opportunities for learning. The subsystem that deals with predator avoidance (and perhaps others) is expected to inhibit exploration. The impact of these subsystems on behaviour should correspond to the relative probabilities of these beneficial and harmful effects; in other words, they should reflect the affordance probabilities of the animal's environment. Note that we are not proposing any particular classification of subsystems; we are simply exploring the utility of this general approach. Note also that the subsystems account is functionally equivalent to the Curiosity-Fear theory, but differs from it in that Li invokes an explanation at what biologists refer to as the level of ultimate causation. Animals may use emotions such as curiosity and fear to manifest ultimate goals in their behaviour, but we do not address this issue[2].

---

[1] Others (MacArthur and Piankam [1966]; Roithlat [1985]) refer to this process as "optimal decision making-. This term is somewhat inappropriate because the behaviour need not be optimal.

This project consists of two parts: empirical work, and simulation. In the laboratory, we compared the behaviour of two groups of banded killifish (Fundulus diaphanus) from two natural populations in a novel environment. Killifish are cyprinadontids found in the weedy, sandy shallows (less than four meters in depth) of lakes and ponds. The Lake Opinicon population is syrnpatric (cohabits the lake) with piscine predators; the Fish Lake population is allopatric from (does not cohabit the lake with) predators. In previous studies, Lake Opinicon killifish maintained a significantly greater mean horizontal distance, and were more likely to move to shallow water in the presence of a largemouth bass, than killifish from Fish Lake (Dunlop, 1987). It seems reasonable to expect that in addition to exhibiting differences in behaviour in the presence of a predator, these populations might show differences in behaviour that could affect the likelihood of encountering a predator in the first place. These behavioural differences would be expected to reflect the relative probability of encountering a predator in each of the two lakes. Thus, differences in predation between the two lakes could give rise to intraspecific variation in behaviour in the absence, as well as in the presence, of a predator.

Since the two lakes do not differ, as far as we know, in the probability of yielding something beneficial such as food, but do differ in the probability of yielding a predator, it was hypothesized that while the extent to which certain subsystems Excite exploration is comparable in the two populations, the extent to which the predator avoidance subsystem inhibits exploration is greater in Lake Opinicon killifish than in Fish Lake killifish.

The goal of the simulation experiments was twofold. First, I wished to confirm that the above model of exploration can generate the general pattern of results observed in the laboratory. Second, I wished to isolate the variable(s) that give rise to differences in the behaviour of the model that parallel the behavioural differences observed between the two killifish populations.

---

[2] For discussion of how exploration satisfies curiosity, see (Berlyne, 1950). For discussion of exploration and fear, see (Russel, 1973; Halliday, 1966). See also (Fowler, 1967) on exploration and reduction of boredom, and (Leuba, 1962) on exploration as a means of maintaining a constant level of arousal.

## 2. LABORATORY EXPERIMENTS

### 2.1 SUBJECTS AND COLLECTING SITES

Nonbreeding killifish were collected from Fish Lake, Prince Edward County, Ontario, Canada, and Lake Opinicon, Leeds County, Ontario, Canada, between May, 1987 and January, 1988. Fish Lake (440 00' N, 770 10' W) is a small, irregularly-shaped lake (Terasmae & Mirynech, 1964). Fish Lake killifish may have been isolated from other killifish populations since the lake formed, approximately 10,000 years ago. A less likely possibility is that they were introduced to the lake more recently by human intervention. The fish community of this lake has been sampled since 1965 and has not been found to contain piscivorous fish (A. Kama pers. comm.) It is likely that Fish Lake is too shallow to support large fish, due to winterkill. Thus, Fish Lake killifish have not been exposed to piscine predators.

Lake Opinicon (44° 30' N, 76° 30' W) is a large shallow lake that is part of the Rideau Lakes canal system (Turnbull, 1975). Lake Opinicon killifish are sympatric with a number of piscine predators (e.g. largemouth bass, Micropterus salrnoides northern pike, Esox lucius and yellow perch, Perca flavescens).

Killifish are a schooling species (Scott and Grossman, 1973), and isolating individual fish disrupts normal behaviour. Therefore, experiments were conducted on groups rather than individuals. Whether or not there are differences in home range size and social group size between the two populations is unknown. Both populations feed on small crustaceans and insect larvae (Keast and Webb, 1966; Kea.st, unpublished data).

Fish from Lake Opinicon were caught with a seine net, and fish from Fish Lake were caught either with a seine net or a funnel trap. Only fish between five and six centimeters in total length (which corresponds to approximately one year of age, Scott & Grossman, 1973) were used. Fish were transported to Queens University, Kingston, in large plastic pails containing garbage bags filled with oxygenated lake water.

### 2.2 HOUSING CONDITIONS

In the laboratory, fish were housed in groups of 25-50 in flow-through aquaria outfitted with immersion heaters. Water temperature was maintained at 20-23° C. When the temperature of the lake water was lower than this, the temperature of the tanks was increased by approximately one degree per day until the desired temperature was reached, thus minimizing thermal stress. Fish were fed Murex freeze-dried ocean plankton once daily. The light-dark cycle was 12L:12D.

### 2.3 EXPERIMENTAL APPARATUS

Experiments were conducted in a tank (120 x 45 x 45.5 cm deep) divided by sheets of opaque white plexiglass into four chambers: a holding chamber (HC) (30 x 15 cm), an experimental chamber (EC) (120 x 30 cm), and two unused chambers (UC) (45 x 15 cm), (Figure 1). The sheet dividing the experimental chamber from the holding chamber had a door (12 cm wide x 15 cm high). Water depth was 30 cm.

The tank was illuminated by three 95 Watt incandescent lighting fixtures placed 90 cm from the top of the tank. A video camera was positioned directly above the tank, 116 cm from the top of the tank, and was connected to a video camera recorder. The holding chamber contained an airstone, and one of the unused chambers contained a heater.

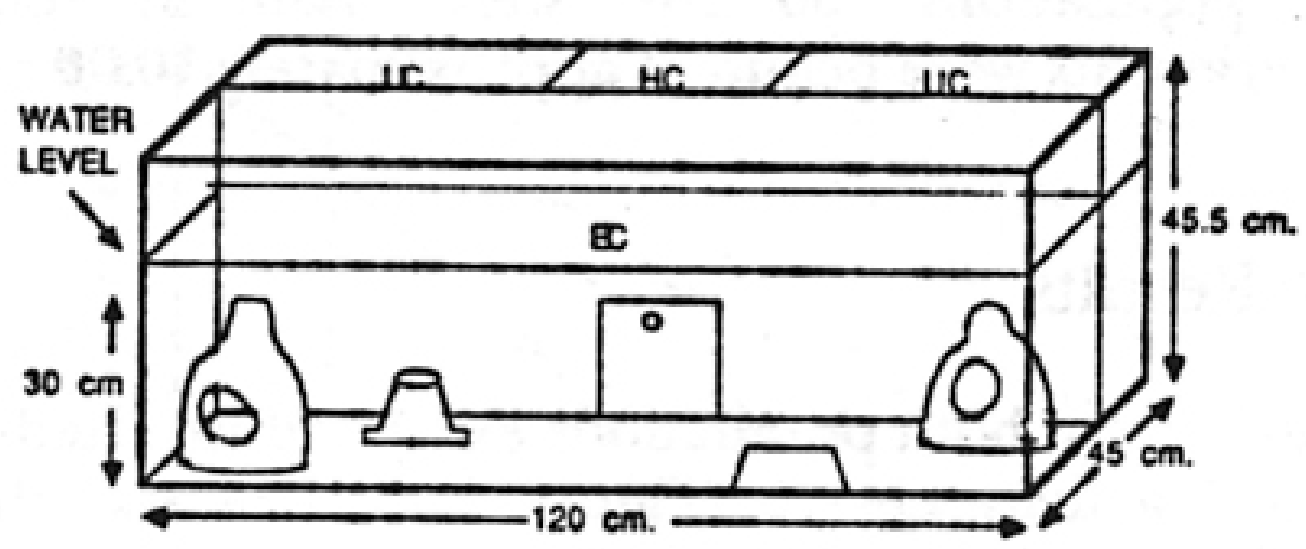

Figure 1. Tank in which killifish were exposed to a novel environment_ EC, Experimental Chamber; IIC, Holding Chamber; UC, Unused Chamber.

The glass along the sides of the experimental chamber was covered with white plastic, to which four panels of wallpaper of various colors and designs (20 x 15 cm) were attached. The wallpaper designs did not resemble anything the fish would encounter in their natural environment. The experimental chamber contained two empty plastic bottles and a small red clay flower pot. Each 01 these items had large holes cut out of them. Two slabs of plexiglass (approximately 3 x 10 cm) were glued to the sides of the tank, and another to the floor, with silicon cement. A fourth plexiglass slat (approximately 15 x 15 cm) was leaned against the side of the tank and at an angle of approximately 45 degrees, and fixed into place with silicon cement The purpose of these items was to ensure that the tank was a novel environment for the fish, and to invite exploration of this environment.

The experimental chamber was covered with a clear plexiglass sheet upon which a numbered grid was drawn. The dimensions of the grid spaces were 10 x 10 cm. Because of refraction and perspective from the camera, equally-sized grid spaces did not correspond to grid columns of equal volume. However, since it was not the use of space per se, but the comparative use of space between the two populations, that was of interest, small differences in the volumes of grid columns had little effect on the conclusions drawn.

### 2.4 EXPERIMENTAL PROTOCOL

Five fish from one of the populations were placed in the holding chamber. 24 hours later, the door to the experimental chamber was opened. The behaviour of the fish was recorded on the VCR for 20 minutes, starting when the door was opened. *Whole-body activity* — movement of the entire body through space — was measured as the sum total number of grid line crossings by all five fish per minute. Three trials were conducted for each of the two populations; 30 fish were used in total. Experiments were begun at approximately 10:00 am.

### 2.5 RESULTS

Fish from both populations exhibited inverted-U-shaped activity curves (Figure 2), The fact that activity over time was nonuniform is strong evidence that it reflected the novelty of the environment, and it is therefore not unfitting to refer to it as exploration.

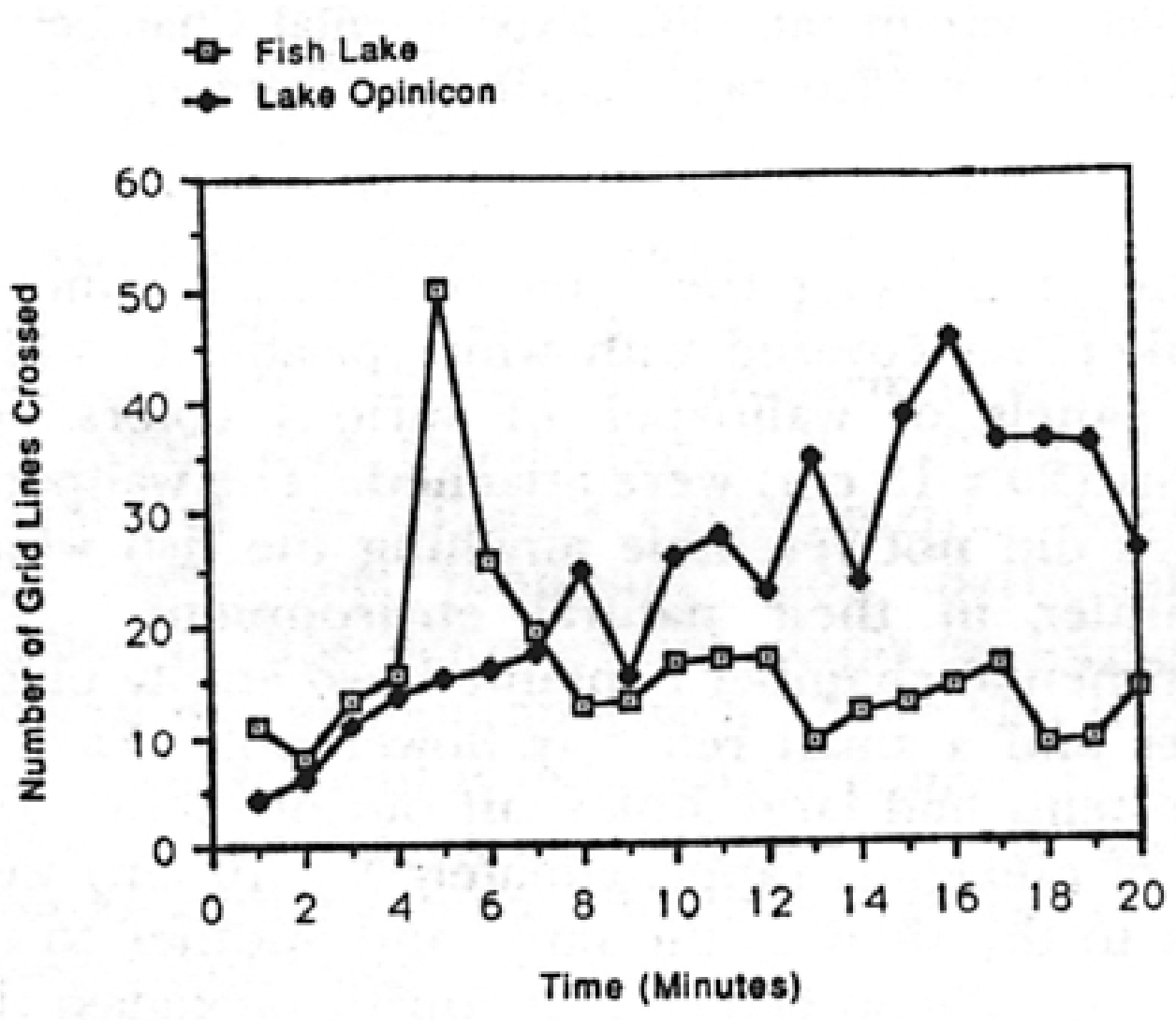

Figure 2. Exploration curves for Fish Lake and Lake Opinicon killifish, plotted as the total number of grid lines crossed by all five fish per minute of exposure to a novel environment.

The shapes of the two curves differ. The activity of Fish Lake killifish rose sharply, reaching a maximum value (51.7 grid spaces/minute) after five minutes' of exposure to the novel environment. By eight minutes, it had returned to the initial level (around 12 grid spaces/minute), and it did not deviate from this by more than five grid spaces/minute for the remainder of the experiment. 335 grid lines were crossed in total.

The initial activity level of Lake Opinicon was 6.3 grid spaces/minute. Activity rose much more gradually and erratically, reaching a maximum value (45.3 grid spaces/minute) after 16 minutes of exposure to the novel environment. It subsequently declined, but still had not return to the initial level by the end of the twenty minute trial. The inverted U-shape of this curve was indicated by the near significance of a quadratic term in regression ($p = .057$). The total number of grid lines crossed was 497.

Fish Lake killifish tended to school (swim within two body widths of one another, oriented in the the same direction) more frequently than Lake Opinicon killifish. Fish Lake fish tended to explore the novel environment "region by region", and generally maintained a fairly constant speed. In contrast, Lake Opinicon fish darted quickly from one spot to another, with prolonged periods of immobility interspersed between the periods of rapid movement.

### 2.6 DISCUSSION

In sum, fish from both Fish Lake and Lake Opinicon exhibited inverted-U-shaped exploration curves, and the major differences between the two curves were: (1) the slope of the Fish Lake curve was greater, (2) it peaked earlier, (3) its peak was higher, and (4) the total amount of exploration was greater for Lake Opinicon fish (i.e. the area under the Lake Opinicon curve is greater). It is not possible to prove conclusively that these differences are related to the difference in threat of predation. Nevertheless, they are consistent with the following theoretical model:

When exploration of a novel environment does not lead to the discovery of affect-laden objects such as food or predators, the animal's (not necessarily conscious) perception of the probability that the novel environment will yield such objects decreases. Thus, exploratory behaviour has a negative feedback effect on both subsystems that excite it and subsystems that inhibit it. There are two additional reasons to expect that exploratory behaviour has a negative feedback effect on excitatory subsystems. First, as exploration proceeds, even as-yet-unexplored space might become less enticing, and be perceived as less likely to yield opportunity for learning, because of its similarity to the space just explored. Second, exploration can lead to fatigue.

For these reasons, were it possible to measure and graph the time-course of the extent to which exploration is (1) excited and (2) inhibited as an animal explores, we would expect to get two negatively-sloped curves. Inhibition should additionally decrease as a function of the amount of Lime spent in the novel environment, regardless of the extent to which the environment is explored. This follows from the nature of ecosystems. Harmful things — predators — are relatively likely to come to you; beneficial things — prey — are relatively less likely to come your way. Since predators are not stationary, time spent immobile in a novel environment without encountering predators provides evidence that the environment does not contain predators, but does not tell an animal much about whether the environment contains prey. It follows that activation of the inhibitory subsystem decreases both with exploration and with time (independent of exploration), while activation of the excitatory subsystem decreases with exploration only. Therefore, unless (1) exploration has less feedback effect on the inhibitory subsystem than the excitatory subsystem, and the difference between the two is greater than the effect of the inhibition-specific time factor, it follows that (2) the slope of the inhibition curve is greater that of the excitation curve. As it turns out, the second condition is necessary in order for exploratory behaviour to get going when it does not begin as soon as the animal is exposed to the novelty (this will be elaborated upon in Section 3.1).

We propose the following explanation for the inverted U-shaped exploration curve. Upon initial exposure to a novel environment, if subsystems that inhibit exploration are more active than those that excite exploration, the animal is initially immobile. Even if this is not the case, inhibition is still initially high, and exploration is minimal. As time passes and novelty diminishes, excitation decreases. However, assuming the novel stimuli were not harmful, inhibition decreases more rapidly than excitation, and the net effect is an increase in exploration. Exploration peaks when the difference between excitation and inhibition of exploration reaches a maximum. When the animal begins to perceive the environment as a relatively safe place, inhibition reaches a minimum, and since excitation continues to wane, exploration decreases. It eventually bottoms out when the animal becomes habituated to the environment (i.e. the environment has lost its novelty). This model is illustrated in Figure 3.

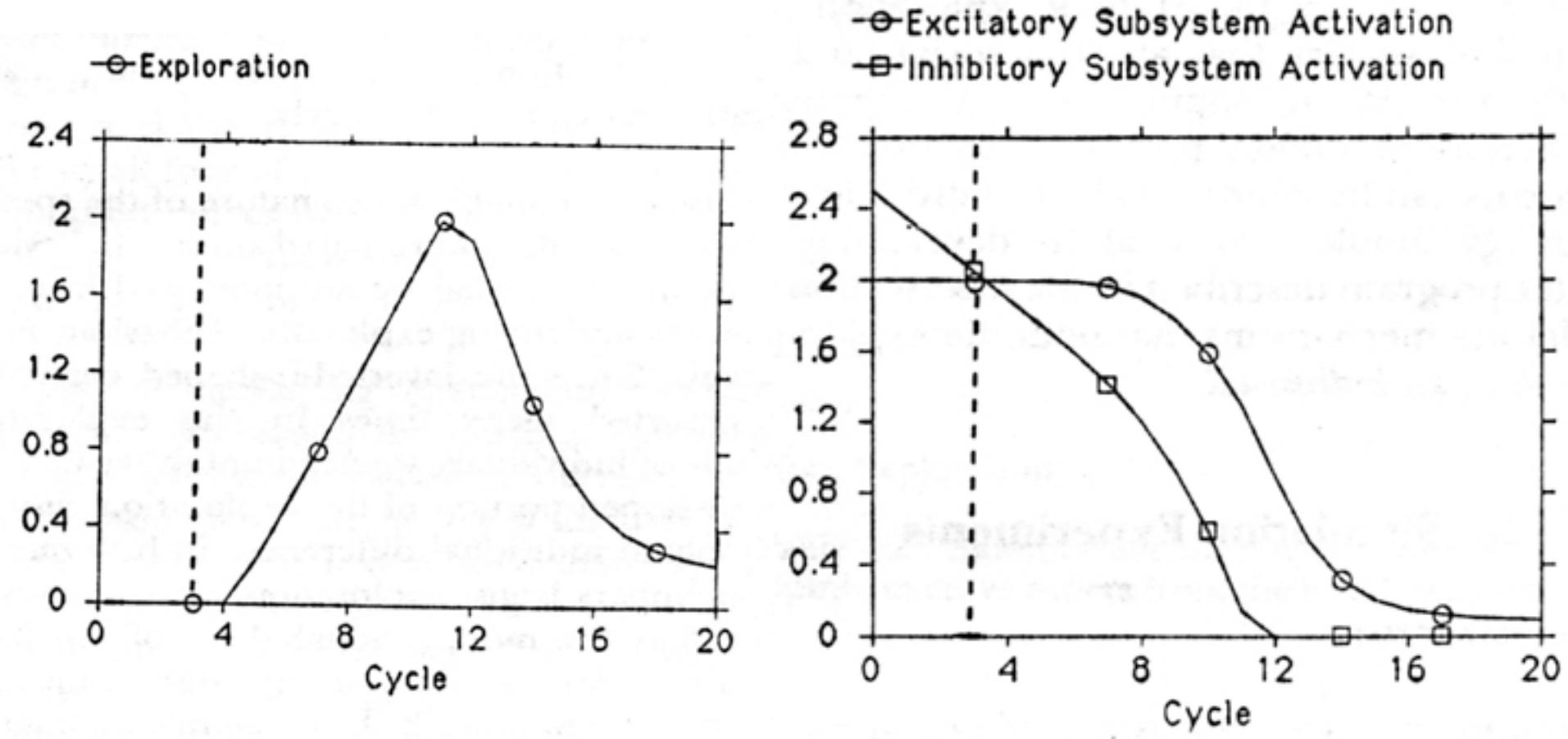

Figure 3. (a) An inverted-U-shaped exploration curve. (b) Activations of the corresponding excitatory and inhibitory motivational subsystems. Dotted line indicates point in time at which excitation surpasses inhibition, enabling exploratory behaviour to begin. Both curves were produced by the computer program described in Section 3. Initial activation of the inhibitory subsystem is 2.5. All other parameters took the default values given in Section 3.1.

The fact that the total number of grid lines crossed was greater for Lake Opinicon fish than for Fish Lake fish is somewhat surprising. Since, given sufficient time, Lake Opinicon killifish engaged in more exploration than Fish Lake killifish, the extent to which exploration is excited by subsystems that could benefit from it is at least as great for Lake Opinicon killifish as for Fish Lake killifish. Since Lake Opinicon killifish took longer to begin exploring, results are consistent with the hypothesis that the populations differ only in the extent to which exploratory behaviour is initially inhibited. The fact that the slope of the Fish Lake curve is steeper than the Lake Opinicon curve is also consistent with this theory. For Fish Lake killifish, inhibition is virtually nonexistent, and exploration increases more quickly. Since excitation is inversely related to the animal's perception of the amount of novelty left to explore, when exploration proceeds quickly, excitation diminishes faster, and exploration dies out more quickly.

Because of the highly social nature of the species used, laboratory data were based on the behaviour of groups; however, what we are interested in is the mechanisms underlying exploratory behaviour in an individual. Since the inverted-U-shaped curve has been reported many times in the exploratory behaviour of individuals, we need not worry that the positively-sloped portion of the exploration curve is due simply to individual differences in how quickly group members begin exploration. It is, however, possible that the more gradual slope of the Lake Opinicon curve is due to greater individual variability in how quickly Lake Opinicon killifish increase and decrease their rate of exploration. We hope to find species that are less social, and for which there exist two populations that experience very different predation pressures, so that similar experiments can be conducted with individuals. To keep things simple, our goal in developing the computer program described in the next section was to model the mechanisms that underlie exploratory behaviour in an individual.

## 3. SIMULATION EXPERIMENTS

### 3.1 ARCHITECTURE

A computer model consisting of two subsystem units and a behavioural output unit was constructed in Common Lisp. One subsystem, the excitatory subsystem is linked by a positive weight to the output unit; it can be thought of as corresponding to a subsystem that could benefit by exploration, such as the subsystem concerned with finding food. The other subsystem inhibits the output unit; it can be thought of as the subsystem that takes care of predator avoidance. Both subsystems are in a negative feedback relationship with the output unit; when activation of the output unit is high enough to cause exploration to surpass a threshold, activation of the subsystems decreases, and consequently their impact on the output unit decreases. Activation of the inhibitory subsystem also decreases as a function of time, independent of exploration, by a constant that will be referred to as the time-factor. The time-factor is a vital component of the model when initial exploration is zero; without it, activation of neither subsystem changes, so the impact of the excitatory subsystem on the output unit never overtakes that of the inhibitory subsystem. The time factor ensures that sooner or later excitation surpasses inhibition and exploration therefore gets underway, (When initial exploration is not zero, the same effect can be achieved by making the feedback weight to the inhibitory subsystem greater than that to the excitatory subsystem.)

Output for each cycle is either a zero, signifying immobility, or a positive number; the larger the number, the greater the amount of exploration prescribed by the system. The relevant variables are:

$s_1$ = activation of excitatory subsystem

$s_2$ = activation of inhibitory subsystem

$w_1$ = weight from excitatory subsystem to output unit

$w_2$ = weight from inhibitory subsystem to output unit

$w_f$ = feedback weight from output unit to each subsystem

$k_1$ = decay on output unit

$k_2$ = time factor

E = amount of exploration

Exploration was calculated as follows:

$$E_{(t)} = \max \{0, k_1 E_{(t-1)} + w_1 s_1 - w_2 s_2\}$$

New subsystem activation values were calculated as follows:

$$s_{1\,(t)} = \max \{0, s_{1\,(t-1)} - w_f E_{(t-1)}\}$$

$$s_{2\,(t)} = \max \{0, s_{2\,(t-1)} - w_f E_{(t-1)} - k_2\}$$

Note that if there was no exploration during the previous cycle, activation of the exciting subsystem does not decrease, but activation of the inhibiting subsystem does.

Three parameters were varied: (1) initial activation of the exploration-inhibiting subsystem, (2) weight from the exploration-inhibiting subsystem to the behaviour unit, and (3) time factor. These parameters affect the impact of the exploration-inhibiting subsystem on the output unit. Earlier we hypothesized that, since exploration could lead to an encounter with a predator for Lake Opinicon killifish but not for Fish Lake killifish, we might expect that the extent to which the predator avoidance subsystem inhibits exploration is greater in Lake Opin icon killifish than in Fish Lake killifish, If this is the case, these parameters are the ones most likely to be functionally similar to the parameters that give rise to the differences in exploration between Fish Lake killifish and Lake Opinicon killifish. It was hoped that by determining which (if any) of these manipulations give rise to differences in curve shape qualitatively similar to the differences between the Fish Lake and Lake Opinicon curves, we could refine our model of how animals tune the mechanisms underlying exploration in a way that matches the affordance probabilities of their environment.

### 3.2 RESULTS

Plotting the program's output did indeed yield an inverted-U-shaped curve. This outcome was robust across all sets of values for initial parameters tested. Initial values of parameters used for all data presented in this paper are unless stated otherwise as follows:

$s_1 = 2.0$

$s_2 = 2.0$

$w_1 = 1.0$

$w_2 = 1.0$

$w_f = 0.5$

$k_1 = 0.5$

$k_2 = 0.15$

By decreasing the time factor (k2) from 0.36 to 0.05, a steep curve with an early, high peak (i.e. a Fish Lake-type curve) was transformed into a more gradually-sloping curve with a lower, later peak, that subsumed a greater area (i.e. a Lake Opinicon-type curve). (These particular time factor values were chosen because they facilitate comparison of program output with laboratory results.) Thus varying a single parameter results in curves that differ by all four of the features that distinguish the exploration curves of the two populations (Figure 4). Total exploration over 20 cycles was for 13.5 Curve 1 and 16.5 for Curve 2.

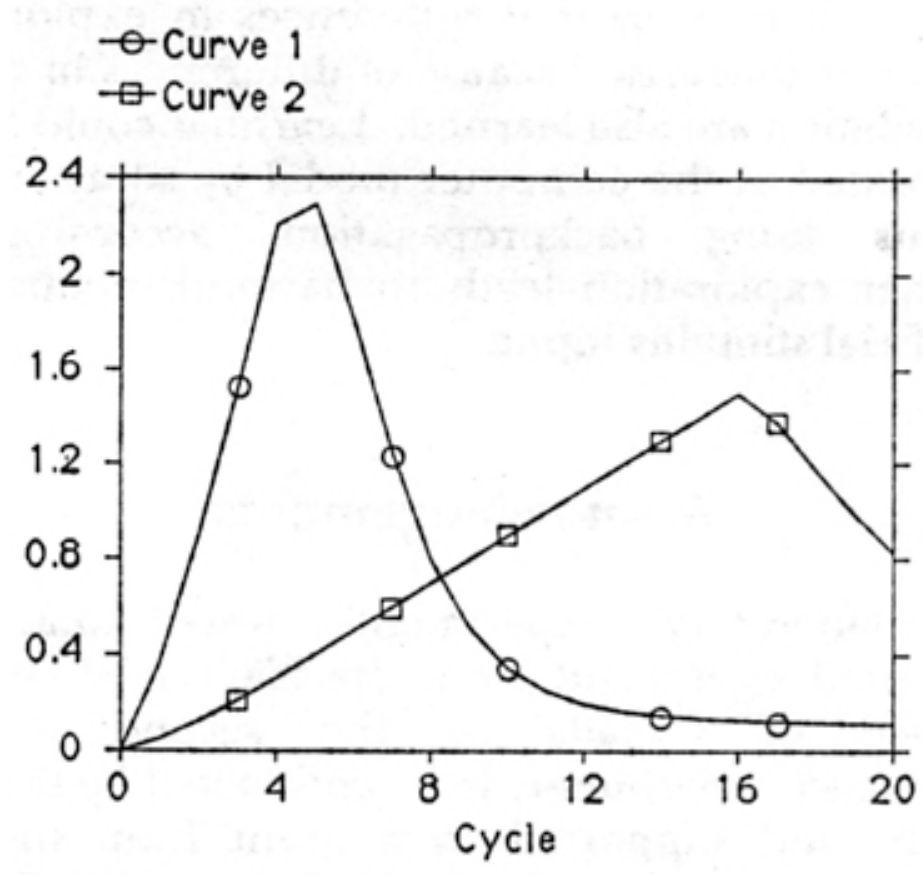

Figure 4. Exploration curves produced by two runs of the computer model. Time factor was 0.36 for Curve 1 (the Fish Lake-type curve), and 0.05 for Curve 2 (the Lake Opinicon-type curve). All other parameters had the same initial values on both runs.

Effects of the varying other parameters were less impressive. Increasing the initial activation of the inhibitory subsystem caused a decrease in peak height, and delayed the timing of the peak, but had no effect on slope. It decreased total exploration. Decreasing initial weight from the inhibitory subsystem produced the same effects. The results of these experiments are summarized in Table 1. A summary of the ways in which the Lake Opinicon curve differs from the Fish Lake curve is presented for comparison.

| | Fish Lake | Lake Opinicon |
|---|---|---|
| Smaller Slope | | x |
| Later Peak | | x |
| Lower Peak | | x |
| More Exploration | | x |

**Table 1a.** Summary of the ways in which the Lake Opinicon curve differs from the Fish Lake curve.

| | Incr. Act. | Incr. Wt. | Decr. TF |
|---|---|---|---|
| Smaller Slope | | | x |
| Later Peak | x | x | x |
| Lower Peak | x | x | x |
| More exploration | | | x |

**Table 1b.** Results of manipulating network variables. Act, activation; Wt, weight; TF, time factor.

### 3.3 DISCUSSION

By tuning one parameter, the program produced a curve which differed from the original along the same four dimensions as differentiate the two killifish curves. This of course does not prove that the mechanisms invoked are similar, even at a course-grained level. Moreover, the specific features that differentiate the killifish exploration curves presented here might not generalize. However, findings fit together nicely with the ecological situation. Lake Opinicon killifish are more likely than Fish Lake killifish to have experienced time-slices in which they didn't encounter predators followed by time-slices in which they did. Therefore, it makes sense that spending a given amount of time in a novel environment without encountering a predator decreases the probability that the environment will afford a predator more for Fish Lake killifish than for Lake Opinicon killifish. The time factor, the parameter that accounted for the four differences in the model that paralleled those in the two populations, provides a means of integrating time-dependent probability assessments in to the equation that determines exploratory behaviour. It is possible that this parameter has a functional equivalent in living brains. In sum, despite limitations, the model implemented in this program has a number of things going for it: (1) It makes sense from an aclaptationist standpoint, (2) It is consistent with the available empirical data, (3) It is computationally feasible, and (4) It seems to be the simplest account that meets the criteria of the first three statements.

## 4. SUMMARY AND CONCLUSIONS

The premise underlying this research is that an animal's behavioural final common path arises from the dynamics of specialized motivational subsystems. Subsystem parameters are tuned in a way that (1) enables the animal to satisfy constraints imposed by both its body and the environment, and (2) takes into account not only the present situation, but the relative probabilities of possible future situations. The time-course of exploration reflects the dynamics of subsystems that could be benefited or harmed by exploration, and these subsystems are tuned to match the probabilities of these possible beneficial or harmful effects.

We have shown that a computer model based on this theory can generate the pattern of results observed in the laboratory. The amount of exploration prescribed by the program is a function of the activation of subsystems that excite and inhibit exploration. Subsystem activation decreases as a function of exploration; activation of the inhibitory subsystem also decreases as a function of time, independent of exploration. When the program is run, and exploration is plotted against time, the result is an inverted-U-shaped curve typical of the time-course of exploration exhibited by animals in a novel environment

The exploration curve from data using killifish that are allopatric with predators was steeper, had a higher, earlier peak, and indicated less total exploration than data using a population sympatric with predators. All four of these differences could be reproduced in the model by tuning a single parameter: the time-dependent component of the decay-rate of the exploration-inhibiting subsystem. It is hoped that, as a package, this set of results is illustrative of how empirical work and simulation can complement one another, and thereby deepen our understanding of the mechanisms that underlie adaptive behaviour.

The question arises as to whether the behavioural differences observed between these two populations are learned — result from 'close calls" with predators or observation of attacks on conspecifics[3] — or whether they are innate. There is evidence for both. The tendency to explore a novel object can be bred for or against in laboratory rodents (Fuller and Thompson, 1978), and reliable differences in the tendency to explore have been shown in different strains of mice (Fink and Reis, 1981). Csanyi (1988) conducted a behaviour-genetic analysis of the paradise fish (Macropodus opercularis), and found that many behavioural traits, including some related to exploration, such as how long fish took to enter a novel environment, were genetically-based or had a strong genetic component However, in another study, Csanyi found that paradise fish that had previously been chased by a pike subsequently attempted to escape when placed in the presence of a pike, whereas naive paradise fish did not (1985). This indicates that, in this species at least, antipredator behaviour can be learned.

[3] Conspecifics are other members of one's own species.

To the extent that antipredator behaviour is learned, it is likely that differences in exploratory behaviour that arise because of differences in threat of predation are also learned. Learning could be in-corporated in the computer model by adjusting the weights using backpropagation, according to whether exploration leads to harmful, neutral, or beneficial stimulus input.

## ACKNOWLEDGEMENTS

The laboratory experiments were financially supported by a grant from the National Research Council of Canada to the second author. Simulation experiments were conducted by the first author, and supported by a grant from Indiana University. Thanks to David Chalmers, Mike Gasser, Stewart Wilson, and Bill Timberlake for helpful comments. Also thanks to Rick Beninger, Rudy Eikelboom, Michael Fox, V. Gotceitas, Dolph Harmson, and Jim Sutcliffe for comments on an earlier draft of section 2.